\documentclass[conference]{IEEEtran}
\usepackage{xcolor}
\usepackage{graphicx} 

\ifCLASSINFOpdf
\else
\fi
\usepackage{amsmath}
\usepackage{amsfonts}
\begin{document}

\title{Quantum Analogues for Two Simple Classical Channels}

\author{\IEEEauthorblockN{Miles Miller-Dickson}
\IEEEauthorblockA{School of Engineering,\\
Brown University\\
Providence, RI 02912\\
{\small miles\_miller-dickson@brown.edu}}
\and
\IEEEauthorblockN{Christopher Rose}
\IEEEauthorblockA{School of Engineering,\\
Brown University\\
Providence, RI 02912\\
{\small christopher\_rose@brown.edu}}
}

\newcommand{\cnote}[1]{{\bf \boldmath \color{red}  {Chris: \em #1}}}
\newcommand{\mnote}[1]{{\bf \boldmath \color{blue} {Miles: \em #1}}}
\newcommand{\be}{\begin{equation}}
\newcommand{\ee}{\end{equation}}

\newcommand{\equat}[1]{equation (\ref{eq:#1})}
\newcommand{\Equat}[1]{Equation (\ref{eq:#1})}

\newcommand{\xv}{{\bf x}}
\newcommand{\yv}{{\bf y}}
\newcommand{\Xmat}{{\bf X}}
\newcommand{\Ymat}{{\bf Y}}

\maketitle

\begin{abstract}
We present some of the peculiar dynamics of two simple sans-entanglement quantum communication channels in a digestible form. Specifically, we contrast the classical gaussian additive channel to its quantum analogue and find that the quantum version features a capacity with interesting time dependence and counterintuitive effects of quantization. We also consider a simple two-level system and comment on the time dependence of its capacity. 
\end{abstract}


% \fi
\IEEEpeerreviewmaketitle

\section{Introduction}
Quantum channels refer to the transmission of classical or quantum information involving the
exchange of quantum states (and sometimes classical states as well). In contrast to classical
channels, quantum channels may make use of entangled particles or superpositions of quantum bits, or qubits, to
transmit information. These channels have an extensive literature describing their information capacity,
their limitations and their advantages over classical transmission \cite{Bennett,Marinescu,Holevo2020,Gyongyosi}. Typically, quantum
channels are described mathematically using density matrices and trace-preserving maps. Here, we present a simplified discussion of two kinds of quantum channels that only assumes
basic knowledge of quantum mechanics. Our aim is to directly compare, where possible, classical
channels to their quantum analogues and to reach an audience who may not be familiar with the density
matrix approach to quantum mechanics.

Quantum mechanics has grown from its esoteric origins to an essential part of pop culture. This
emergence is partly due to enormous non-scientific popularization, but it is also due to recent
explosive development in quantum technologies such as quantum sensing, computation, cryptogrphy,
etc. \cite{MacFarlane,Sigov}. The rise in prevalence of these technologies is bringing with it a broader demand to
appreciate the subtleties of quantum systems. The topic of quantum channels, in particular, has far
reaching implications for the effectiveness of many of these technologies, as they can often be
framed as communication problems. Among these subtleties are the counterintuitive effects of
entanglement. Even before these effects, however, are the more basic effects of
interference (superposition), which are alone already responsible for so much of the ``strangeness'' of the
subject.

In this paper, we consider precisely some of those strange effects in the context of two basic channels. The first is a quantum analogue of the additive Gaussian channel. The second is a
two-level system which provides a comparison to single bit transmission in a classical setting. In section II, we introduce the first of
these quantum channels and compute its capacity. We present the second channel in section III. And
section IV concludes with a discussion on the comparison of these channels to their classical
counterparts. In
an attempt to keep this work self-contained, we include a brief primer in an appendix on the elements of
quantum mechanics that are relevant for this discussion. We also include in the appendix derivations of some results in the main text.

%\cnote{this is actually rather a spectacular opening, Miles.  I could not have done it as well myself.}

\section{Gaussian channel analogue}
Let us consider the classical additive Gaussian channel, framed as
\be
Y = X + Z
\ee
$X\in\mathbb{R}$ is the channel input (sent by Alice), $Y\in\mathbb{R}$ is the channel output
(received by Bob) and $Z\in\mathbb{R}$ is a zero-mean Gaussian random variable, independent of
$X$, interpreted as noise. Alice codes messages as sequences (blocks) of $X$'s, each of which is
corrupted by a sequence of $Z$'s so that Bob receives sequences of $Y$'s. Bob then decodes the
sequences of $Y$'s to recover the messages. It is also usually assumed that $X$ is constrained
( $E[X^2] \propto \mbox{power}$). In the simplest case, the sequence of $Z$'s are independent and identically distributed (i.i.d).

\begin{figure}[h]
\begin{center}
\includegraphics[width=3.0in]{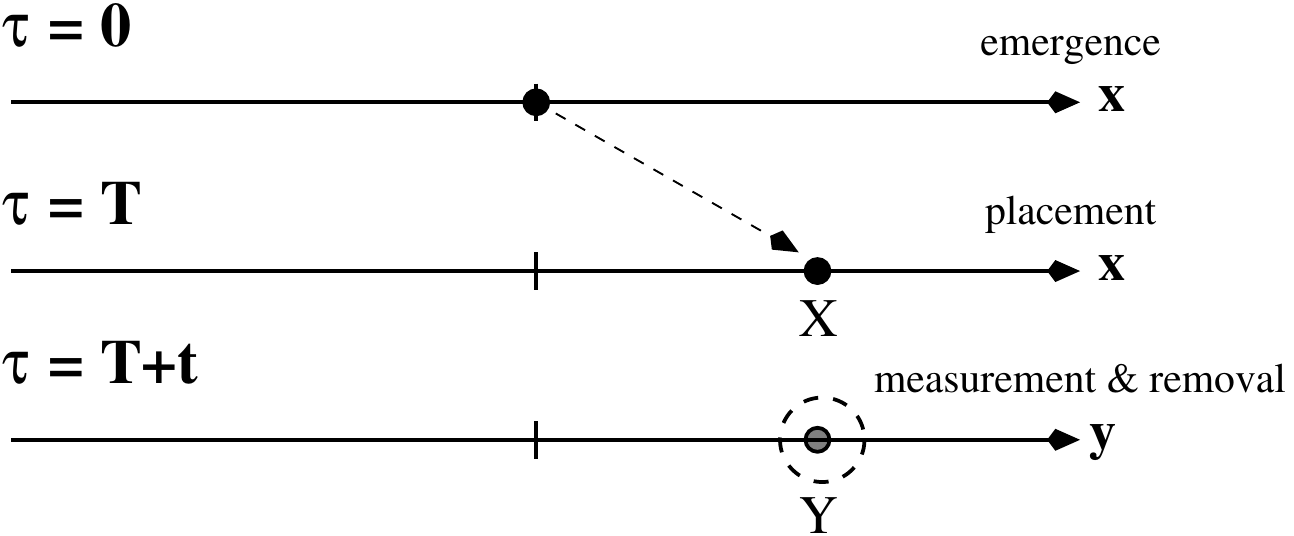}
\end{center}
\caption{{\bf Particle Placement Channel:} a particle emerges at the origin at $\tau = 0$ and is moved by Alice to $X$ by time $\tau = T$.  Bob measures the particle position $Y$ at $\tau = T+t$ and removes it.  This process repeats with period $T+t$ to produce the tuple sequence, $\{X_k,Y_k\}$.}
\label{fig:pebblechannel}
\end{figure}
In our toy quantum channel we use a simple model to describe the physical system. Alice prepares a message
by choosing a sequence of locations along some axis at which to place a particle. Bob then measures
the position of each particle some specified time delay $t$ after it is placed and removes it from the system.
This channel is illustrated in FIGURE~\ref{fig:pebblechannel}.
% Since the particle initially emerges at $x=0$, the energy required for particle placement is proportional to the energy necessary to move a particle from $x=0$ to some $X = x_n$ in a specified amount of time $T$, where $n$ indexes the placement run.  Thus, the minimum average power $P$ expended to place particles goes as,
% \be
% \label{eq:Power}
% P
% =
% \frac{1}{2(T+t)} m \mathbb{E} \left [ \left (\frac{X}{T} \right )^2 \right ]
% \ee
% where $\frac{1}{T+t}$ is the rate at which the communication scheme cycles (particles are placed then measured), $m$ is the particle mass and the $\mathbb{E}[\cdot]$ denotes the expected value over particle placement positions $X$. For later notational simplicity we will write \equat{Power} as,
% \be
% \label{eq:Power2}
% P
% =
% \beta \mathbb{E} \left [ X^2 \right ]
% \ee
% where
% $$
% \beta = \frac{m}{2T^2(T+t)}
% $$

We then consider the simplest model for the quantum dynamics---that of the free particle. To
determine the capacity of this channel, we will compute the probability that Bob measures the
particle at some location $Y$ along the axis some time $t$ after Alice places the particle at
an initial location $X$. This process is repeated for each particle placement.

% Thus, for each placement 
% we have a conditional probability distribution $f_{Y|X}(y|x)$ which represents the probability Bob
% finds the particle at $Y$ given Alice placed it at $X$.  Put another way, $f_{Y|X}(y|x)$ is the
% channel transition function so that owing to the assumption of independent channel uses, the
% transition function for a sequence of channel uses obeys,
% \be
% f_{\Ymat|\Xmat}(\yv|\xv)
% =
% \prod_{n=1}^N
% f_{Y_n|X_n}(y_n|x_n)
% \ee
% where boldface variables denote vectors.

We will assume that Alice can place the particle at some location only up to some finite precision,
limited by the nature of the apparatus. This is effectively the quantization precision that would be
present in practice even for the classical channel. We will characterize this placement precision
using the variance of the initial particle wavefunction constructed by Alice---technically it is the variance of the distribution $|\psi(x)|^2$. A lower wavefunction
variance implies higher precision in Alice's placement. Classically, increasing the quantization
precision has diminishing returns for the capacity \cite{Host}, although for our quantum system we will
find that in addition to the usual diminishing returns, increasing precision can actually {\em
  reduce} the capacity of the channel.

\subsection{Time evolution of the Gaussian particle}
For computational ease, we assume that the placement of a particle by Alice is described by a
Gaussian wavefunction with some variance $\sigma_A^2$. The exact wavefunction that Alice constructs
will not disturb the qualitative aspects of our results, so long as the wavefunction is
localized to some degree about a point.
\be
    \psi(x) = \frac{1}{(2\pi\sigma_A^2)^{1/4}}e^{-(x - x_0)^2 / 4\sigma_A^2}
    \label{eq:wf}
\ee

Notice that $|\psi(x)|^2$ is a Gaussian distribution centered at $x_0$ with variance $\sigma_A^2$. $\psi(x)$ is a purely real-valued expression, but it could have any complex phase factor scaling it, such as $\exp(ig(x))$ where $g(x)$ is some real-valued function of $x$---this factor would not disturb our interpretation of $\psi(x)$ as a particle localized at $x_0$ since the modulus is unaffected by such a factor. The Gaussian choice is particularly convenient because a Gaussian wavefunction will evolve into a Gaussian with expanding variance. 

In the appendix, we calculate the time-evolution of Alice's initial state and take the squared modulus, giving the probability density for locating the particle at point $x$ at time $t$. The result is given by,
\be
    |\psi(x,t)|^2 = \frac{1}{\sqrt{2\pi\Delta_t^2}}\exp{\left(\frac{-(x-x_0)^2}{2\Delta_t^2}\right)}
\ee
where $\Delta_t^2$ is given by (for $t\ge 0$),
\be
\label{eq:Noise}
\Delta_t^2 = \sigma_A^2 + \left(\frac{\hbar t}{2m\sigma_A}\right)^2. 
\ee

Thus, an initial Gaussian evolves into a Gaussian with a variance that grows \textit{quadratically} in time. As a point of contrast, the distribution over the position of a particle experiencing Brownian motion (random walks), has a variance that grows linearly in time \cite{Wang}. As we will see in the next section, if Alice can set the initial variance of the particle's position to be proportional to $t$, the time that Bob waits to make a measurement, it will be possible to recover a noise variance that only grows linearly in $t$, just as in the Brownian case.

\subsection{Capacity of the Quantum Gaussian Channel}
In the previous section, we computed the probability density that Bob will locate the particle at a position some time after Alice placed it along an axis. In what follows, we will use the variable $X$ to refer to Alice's choice in the initial particle location, and we will use the variable $Y$ to denote the r.v. associated with Bob's measurement at some specified time later. Thus in the language of communication theorists, we will write $|\psi(y,t)|^2$ as $f_{Y|X}(y|x)$, the transition probability for our channel (where we have suppressed the dependence of the wavefunction on Alice's initial particle position $x=x_0$). And we use $f_X(x)$ and $f_Y(y)$ to refer to the probability density over Alice's choice of $X$, and the density associated with Bob's measurement, respectively. We will assume that the variance of $X$ is bounded by, 
\be
    \mathbb{E}[X^2] \le P
\ee
for some constant $P$. Note that $P$ has dimensions of ${(length)}^2$. Thus, $\sqrt{P}$ characterizes the length scale over which Alice can place her particle on the axis---we will regard $P$ as a measure of signal strength.

The capacity $C$ is given by maximizing the mutual information over $f_X$, subject to this quadratic constraint on $f_X$, \cite{covernew} 
\be
    C = \max_{f_X, \mathbb{E}[X^2]\le P^\prime} I(X;Y).
\ee
Mutual information can be written as $I(X;Y)=H(Y) - H(Y|X)$, and $H(Y|X) = \int dx f_X(x) H(Y|X=x)$. But since $f_{Y|X}$ is Gaussian with variance $\Delta_t^2$, then $H(Y|X=x) = \frac{1}{2}\ln{2\pi e \Delta_t^2}$, which is independent of $x$ \cite{covernew}. Thus, 
\be
    H(Y|X) = \frac{1}{2}\ln{2\pi e \Delta_t^2}
\ee
So maximizing $I(X;Y)$ is equivalent to maximizing $H(Y)$. 
%(since $\int dx f_X(x) = 1$ \textcolor{blue}{$\leftarrow$ need I bother saying this?})

Now we make an observation about the form that $f_Y(y)$ takes. Since $f_{Y|X}(y|x)$ is Gaussian, then $f_Y(y)$ is given by,
\be
\begin{array}{rcl}
f_Y(y) & = & \int dx f_X(x) f_{Y|X}(y|x) \\
       & = & \frac{1}{\sqrt{2\pi\Delta_t^2}}\int dx f_X(x) e^{-(y - x)^2/2\Delta_t^2}
\end{array}
\ee
which is precisely the same distribution as the classical additive Gaussian channel with a ``noise"
variance $\Delta_t^2$. Thus, for a fixed measurement delay $t$, our quantum analogue is equivalent to
the classical additive Gaussian channel. Really, this quantum version comprises a set of classical
Gaussian channels parameterized by the measurement delay $t$---this follows
directly from the fact that the transition probability $f_{Y|X}$ is Gaussian.

Hence, $H(Y)$ is maximized by a Gaussian with variance $\Delta_t^2+P $, just as it is in the classical case. This means that $H(Y)=\frac{1}{2}\ln{2\pi e(\Delta_t^2 + P)}$, and so the capacity in nats per channel use is given by,
\be
\boxed{
\begin{array}{rcl}
\label{eq:Capacity}
C & = & \frac{1}{2}\ln{2\pi e(\Delta_t^2 + P)} - \frac{1}{2}\ln{2\pi e\Delta_t^2} \\
  & = & \frac{1}{2}\ln{\left(1+\frac{P}{\Delta_t^2}\right)}.
\end{array}
}
\ee
which is superficially the form of the usual Gaussian channel with noise variance $\Delta_t^2$. That this capacity does not exceed the classical value follows from \textit{Holevo's} bound which
essentially states that using quantum states to encode classical information does not buy you any additional
capacity than using the corresponding classical states (at least not in the manner suggested here,
i.e. without making use of entanglement) \cite{Harrow,Holevo2020,Bowen}.

The curious element of this capacity ultimately lies in the ``noise'' power given by \equat{Noise}. Evidently if Bob waits longer to measure the particle position, capacity
decreases, which may not be so different from a classical setting where one could model the noise term
as being time-dependent. Still, a natural model in the classical setting is the Brownian one we
alluded to earlier. But somehow the quantum noise grows even faster than this. 

\begin{figure}[h!]
    \centering
    \includegraphics[scale=0.6]{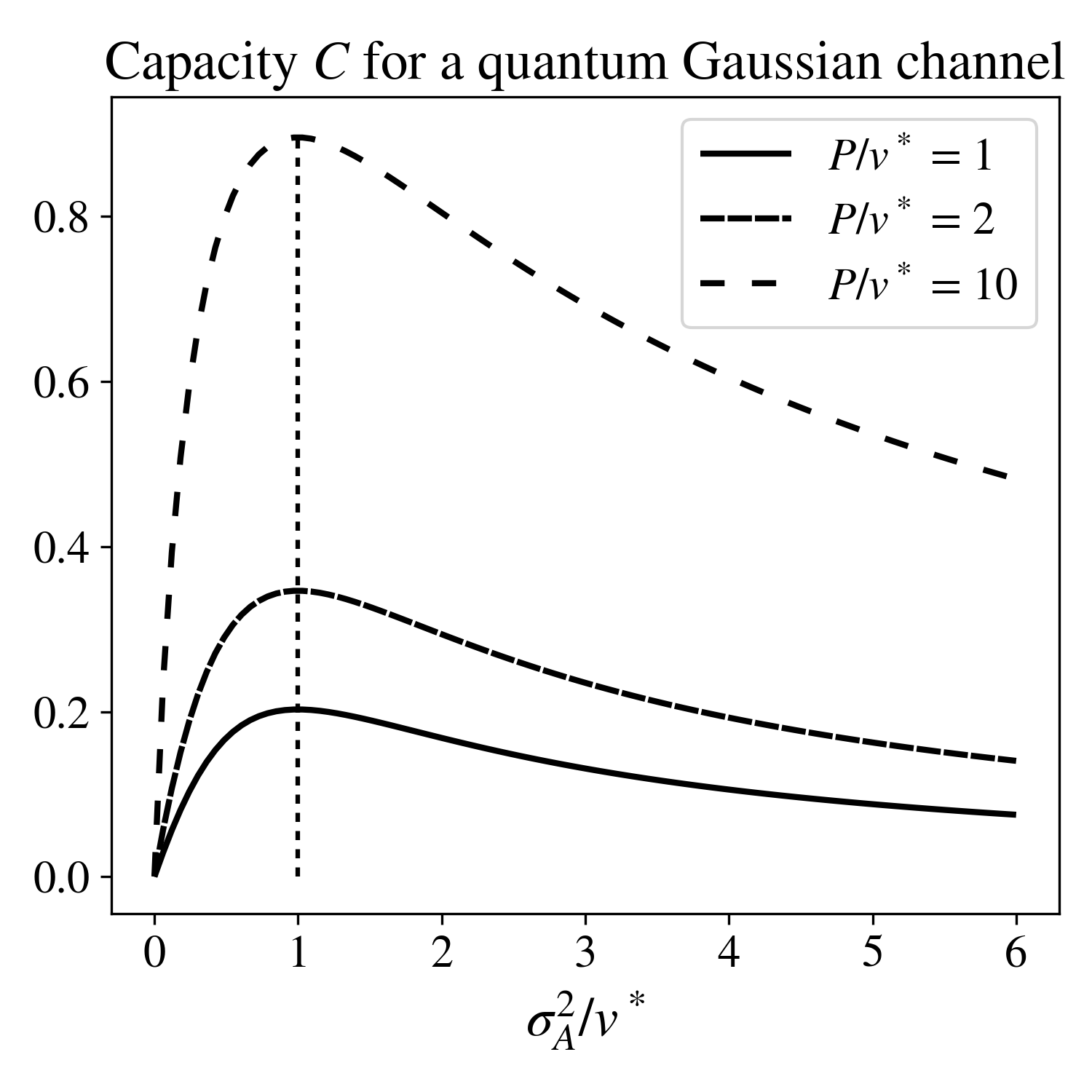}
    \caption{{\bf Capacity of the Quantum Particle Placement Channel:} capacity (in \textit{nats}) is plotted as in \equat{Capacity} with $v^*$ given by \equat{threshold} for different ``signal to noise" ratios $P/v^*$.}
    \label{fig:capacity}
\end{figure}

It is possible, however, for Alice and Bob to optimize the initial choice of precision to reduce the effect of the noise as much as possible. Note that reducing $\sigma_A^2$ can increase the noise power (\equat{Noise}). So in fact for a fixed measurement delay $t$, increased precision on Alice's part (reducing $\sigma_A^2$) can reduce the
capacity. In particular when $\sigma_A^2$ is reduced below a threshold variance $v^*$, given by
\be
\label{eq:threshold}
v^* = \frac{\hbar t}{2m}
\ee
then the capacity will be reduced, as can be seen in FIGURE~\ref{fig:capacity}. Of course the
capacity is also reduced by increasing $\sigma_A^2$, and so $v^*$ represents an optimal placement
precision by Alice to maximize the capacity of this channel, given a predetermined measurement delay $t$. For this choice of $\sigma_A^2$, the noise power is given by $\Delta_t^2 = \hbar t/m$, which grows linearly with the time Bob waits to measure the particle. And so as far as the scaling of the noise in time is concerned, Alice and Bob can only communicate as well as one could in the case of Brownian noise. 

\section{Two-level quantum channel}
\begin{figure}[h!]
    \centering
    \includegraphics[scale=0.4]{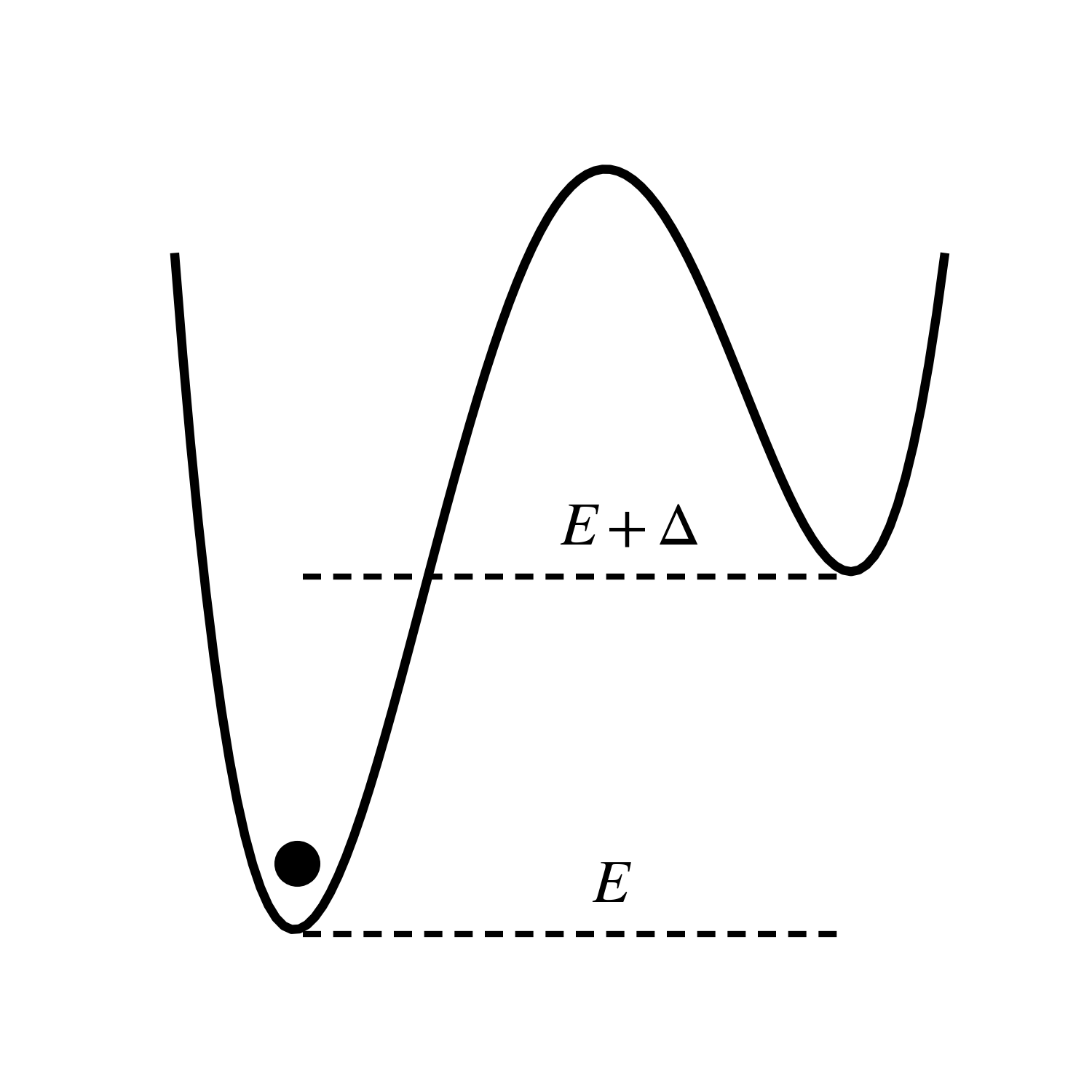}
    \caption{{\bf Two-State Channel:} Alice sets the system in of two states ($|0\rangle$ or $|1 \rangle$) and Bob will measure system state some time $t$ later. This process repeats independently at $t$-second intervals to produce tuples $\{ |\psi_{kt} \rangle, \psi_{(k+1)t} \rangle\}$.}
    \label{fig:doubleWell}
\end{figure}

In a two-level channel, we imagine that Alice places some system in one of two states. Let us generically refer to these states as $|0\rangle$ and $|1\rangle$. These could be the groundstate and first excited state of an atom, spin-$1/2$ states of an atom in a magnetic field, exciton states of quantum dots, etc \cite{Dattagupta,Zrenner}. One can also imagine placing a particle in one of two potential energy wells, as shown in FIGURE~\ref{fig:doubleWell}. Then, some time $t$ later, Bob will measure the state of the particle and either find $|0\rangle$ or $|1\rangle$. In the case of our quantum Gaussian channel, we saw that Alice's initial state evolves over time, giving rise to ``noise" in Bob's measurement---non-zero transition probabilities. Here, we model transitions between the two levels using a parameter $\epsilon\in\mathbb{R}$ in a two-level Hamiltonian of the form,
\be
    \hat H = 
    \begin{pmatrix}
        E & \epsilon \\
        \epsilon & E+\Delta
    \end{pmatrix} .
\ee
When $\epsilon=0$, this corresponds to a two-level system without any transitions between the levels. In that case the $|0\rangle=(1,0)^T$ state has energy $E\in\mathbb{R}$, and the $|1\rangle=(0,1)^T$ has energy $E+\Delta$, $\Delta>0$. For $\epsilon>0$, there is a non-zero probability for transitions between the $|0\rangle$ and $|1\rangle$ states, which is sometimes referred to as \textit{tunneling} between the two states. 

\subsection{Time evolution of the two-level system}
Here we assume Alice can set her initial state as,
\be
    |\psi_0\rangle=\sqrt{1-p}|0\rangle + \sqrt{p}|1\rangle
    \label{eq:2levelState}
\ee
either with $p$ close to $0$ (so that $|\psi_0\rangle$ is close to $|0\rangle$) or with $p$ close to $1$ (so that $|\psi_0\rangle$ is close to $|1\rangle$). Here $p$ gives the initial probability of a measurement resulting in the state $|1\rangle$, and so $1-p$ is the probability of measuring the state $|0\rangle$. Hence, we can use $p(1-p)$ as a measure of the variance for the corresponding distribution, $|\langle z | \psi_0\rangle|^2$, where $z\in\{0,1\}$ (Bernoulli distribution). $p(1-p)$ plays the role here that $\sigma_A^2$ played for the Gaussian channel in section II, and similarly we will assume that it depends on the fidelity of the equipment Alice uses to prepare her state. 

In the appendix we compute the time evolution of this state. Taking the modulus squared of each coefficient gives,
\be
    \label{eq:tProb2level}
    |\langle z | \psi_t\rangle|^2 = \frac{1}{a^2}
    \begin{pmatrix}
        \zeta_p\cos(\frac{2at}{\hbar}) + a^2(1-p)-\zeta_p  \\
        - \zeta_p\cos(\frac{2at}{\hbar}) + a^2p+\zeta_p
    \end{pmatrix}_z,
\ee
where, 
\be
    \zeta_p = \frac{\epsilon}{2}\left(\epsilon(1-2p)+2 b\sqrt{p(1-p)}\right),
\ee
for $z\in\{0,1\}$, representing the two states, and $a = \sqrt{\Delta^2 + 4\epsilon^2}/2$. 

Hence, we see that Bob's likelihood of measuring either $|0\rangle$ or $|1\rangle$ oscillates in time with a frequency given by $f = a/\pi\hbar$. This will mean that the capacity of this two-level channel will also oscillate with that frequency. 

\subsection{Two-level channel capacity}
As before the channel capacity is computed by maximizing the mutual information $I(X;Y)$. 
We will not need to suppose any quadratic constraints on $X$, as Alice only prepares one of two states (not one out of infinity). In this case we assume that Alice will set $|\psi_0\rangle$ in one of two states: either $p=r_0$, where $r_0$ is such that $0\le r_0\ll 1$, or $p=1-r_0\approx 1$. In other words, Alice can bias her state towards the $|0\rangle$ state (setting $p=r_0$) or towards the $|1\rangle$ state (setting $p=1-r_0$). So we can treat $r_0(1-r_0)$ as a measure of the variance associated with Alice's preparation, and as before low variance corresponds to high precision on Alice's part. 

We treat the variable $X$ as a binary random variable taking values in $\{0,1\}$ corresponding to the choices $p=r_0$ and $p=1-r_0$ respectively. $Y$ also takes values in $\{0,1\}$, corresponding to the outcomes that Bob could measure after a delay $t$. Equation~(\ref{eq:tProb2level}) gives the transition probability $p_{Y|X}(y|x) = |\langle y | \psi_t\rangle|^2$ (with $x,y\in\{0,1\}$). These values can be used to compute the mutual information. Unlike for the Gaussian channel, the capacity here can be shown to be monotonically decreasing by increasing $r_0(1-r_0)$, the variance of Alice's starting state, and so Alice can always do better to increase her preparation precision for this channel. 

\begin{figure}[h!]
    \centering
    \includegraphics[scale=0.5]{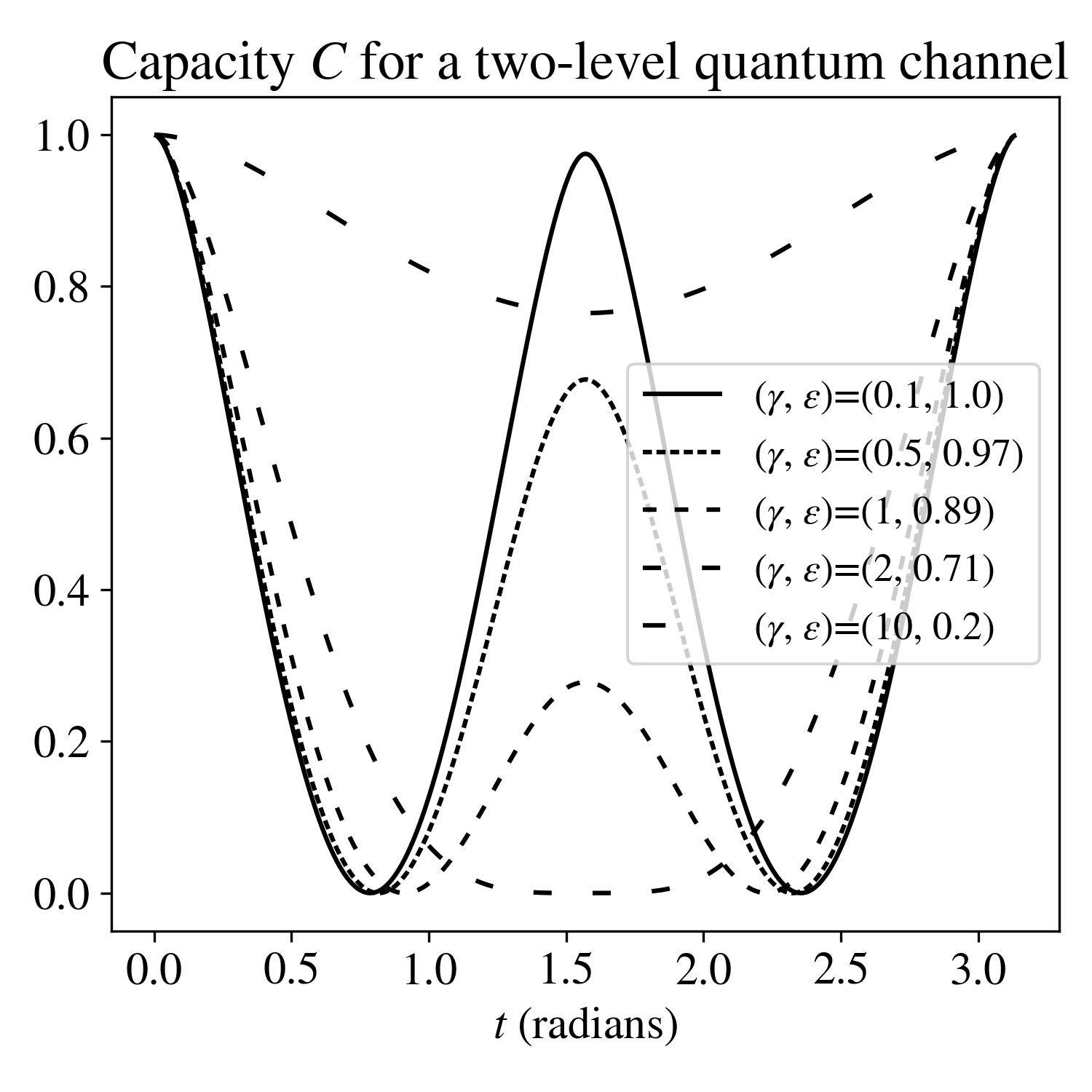}
    \caption{{\bf Capacity  of the Two-Level Channel Over the Course of One Full Cycle:}  capacity in bits is shown for several values of $\gamma=\Delta/\epsilon$, and we set $\epsilon=2/\sqrt{\gamma^2 + 4}$ so that each case has the same frequency for comparison in the same time window.}
    \label{fig:2levelCapacity}
\end{figure}

What is interesting about the capacity in this case is that it oscillates in time. In FIGURE~\ref{fig:2levelCapacity}, we show the time dependence of the capacity over one full cycle. The transition probabilities can be parameterized by $\epsilon$ and by the ratio $\gamma=\Delta/\epsilon$. We choose several values of $\gamma$ and set $\epsilon=2/\sqrt{\gamma^2 + 4}$ in order to fix the frequency and compare capacities within the same time window. 
Depending on the ratio $\gamma$ and the value $\epsilon$, Bob will be able to extract more information from the channel by choosing to measure the state at the right time. In this example, Bob always does best to measure at delay times that are integer multiples of the period of oscillation, $T_0$, given by,
\be
    T_0 = \frac{2\pi\hbar}{\sqrt{\Delta^2 + 4\epsilon^2}}.
    \label{eq:T0}
\ee
In this sense, measurement times must be synchronized with the channel to maximize information transfer.

\section{Discussion}
We have presented two simple quantum models for transmitting classical information, one involving a continuous set of information states, and one involving a discrete set. In this sense these examples live on opposite ends of a discussion on the transmission of information. Our aim was to keep our analysis as self-contained as feasible and thus only assumes very limited prior knowledge of quantum mechanics. The channels we discussed involved preparing a quantum state in a fashion that biased it toward some classical state. Later, the system could be measured and we used this to construct a channel between Alice (preparation) and Bob (measurement). This setup does not make use of entanglement in any manner and yet still provides a demonstration of some of the counterintuitive effects of using quantum states to encode classical information.

In the case of the Gaussian channel analogue, we found that increasing Alice's preparation precision only increased the capacity up to a point. Beyond this threshold, preparing a state (localizing a particle) with any more precision will reduce the capacity of the channel. We saw that this tradeoff was due to the quantum version of noise that is introduced by ordinary time evolution. In particular, a particle that is increasingly localized has a wavefunction that expands more rapidly, leading to increased uncertainty in Bob's measurement, and a reduced capacity.

As an example, if Alice and Bob make use of a particle with mass on the order of $\sim 10^{-27}$ kg (atomic scale), and Bob waits to make a measurement on the order of every second or so, the optimal precision of the particle is confined to a region on the order of $\sqrt{\hbar t/ m}\sim 10^{-4}$ m. If Alice prepares the particle with more precision than this, the variance of Bob's measurement will be increased away from its minimum, and the capacity will be reduced. Any less precision and Alice and Bob will loose out on information contained in the position. 

\begin{figure}[h!]
    \centering
    \includegraphics[scale=0.6]{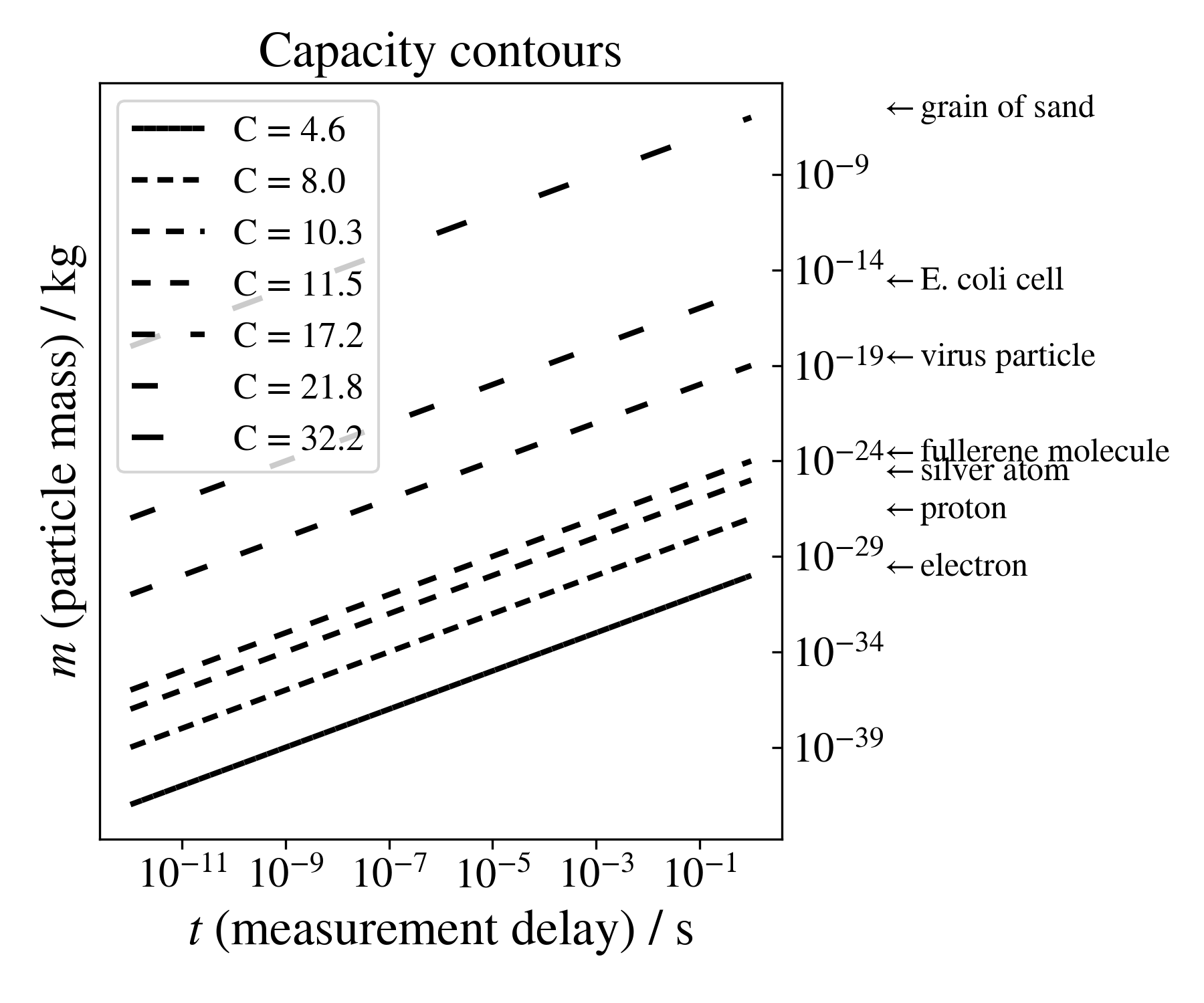}
    \caption{Contours of the capacity $C$ (in \textit{nats}) for the quantum gaussian channel plotted for various particle masses, $m$, and measurement delays, $t$. We set $P=1$m$^2$. Markers indicate the mass values and corresponding capacity values for $t=1s$ measurement delays.}
    \label{fig:massTimeCapacity}
\end{figure}

We see that the variance in Bob's measurement varies inversely with the particle mass. Thus, the more one tries to localize a particle the faster it escapes, but heavier particles escape more slowly. This effect is a consequence of the Heisenberg uncertainty principle---$(\Delta x)^2(\Delta p)^2 \ge \hbar^2/4$, where $\Delta x$ and $\Delta p$ give the standard deviation over possible measurements of the particle's position and momentum, respectively \cite{Sakurai}. If Alice and Bob make use of a more macroscopic particle, say with mass of order $\sim 10^{-6}$ kg, the optimal precision is of order $\sim 10^{-14}$ m, well within the typical length scale of a hydrogen atom ($\sim 10^{-11}$ m). The information content is more stable when stored in more massive particles as the measurement variance expands more slowly. As a result Alice and Bob can increase the capacity of their channel with increased particle precision. The capacity cannot be increased indefinitely however. As the mass continues to grow, the optimal precision length continues to shrink past the subatomic scale. Under these conditions, it becomes increasingly infeasible for Alice to prepare states with such precision, fundamentally limiting the capacity which could be achieved by such a communication scheme. In FIGURE~\ref{fig:massTimeCapacity} we present a comparison of the capacity across various optimal length scales and measurement delays, ranging from the scale of an electron mass ($\sim 10^{-31}$ kg) to that of a grain of sand ($\sim 10^{-6}$ kg).

% In principle, the preparation of the particle's precision is limited by the energy scale available to Alice, governed by the relation $\Delta E\sim \hbar^2/8m\Delta x^2$, where $\Delta E\sim (\Delta p)^2/2m$ gives the standard deviation of possible energy measurements for a particle localized to a region $\Delta x$---another application of the uncertainty principle. This energy range represents the scale over which the particle's energy could likely be measured, reaching increasingly large values as $\Delta x$ is reduced. Alice, however, requires a correspondingly large energy source to \textit{prepare} such a precise state, and so such a communication scheme is ultimately limited by energetic constraints. 

It is interesting to note that quantum mechanics reduces to classical physics in the appropriate limit. A convenient shortcut of describing this limit treats $\hbar$ as an increasingly negligible parameter. In particular, physicists will sometimes ``take the limit" as $\hbar$ tends to zero, which is equivalent to considering the case when various quantities of interest become large compared with $\hbar\sim10^{-34}$ joule-seconds. In this limit, our threshold value $v^*\propto \hbar$ vanishes, recovering the classical notion that one can continue to increase the capacity of the channel by increasing the preparation precision ad infinitum. 

In the two-level channel, on the other hand, we found that the capacity can always be increased by increasing the precision of Alice's preparation. What was interesting in this case was that the capacity is an oscillating function of time, leading to the effect that measurements should be timed according to the frequency of the channel to achieve optimal capacity. Again considering the classical limit, \equat{T0} shows that the time between peaks in the capacity decreases as $\hbar\rightarrow 0$. In which case, it will become increasingly difficult for the receiver to maintain synchronization with the sender and the capacity will average out. As a recent example, quantum dot technology is being pursued as a two-level model for use in quantum computation \cite{Zrenner}. In that context, oscillations occur on the order of hundreds of picoseconds, making it currently unfeasible with such technology to time measurements to integer multiples of the period. In the case of NMR, however, the relaxation time scale is on the order of seconds \cite{NMR1,NMR2}. Regardless of the system, if the energies $\Delta$ and $\epsilon$ could be tuned to be sufficiently small, the receiver can have more time to maintain synchrony with the channel. In this case, improved information transfer relies on Alice and Bob coordinating to maintain a kind of resonance with their two-level channel. 

Lastly, both of our toy quantum systems feature a fundamental obstacle to their use as communication methods. By their nature, both systems are subject to quantum fluctuations whose inherent indeterminacy lead to unavoidable losses in capacity. In the Gaussian case, this was due to the uncertainty principle that trades position and momentum measurement certainty. In the two-level case, this was the presence of a non-zero tunneling rate between the two configurations. These considerations suggest that making use of quantum systems, at least in the method argued here, is fundamentally limited for storing and communicating classical information. At its most basic level, however, any storage / communication system behaves quantum mechanically. It is only in an aggregated form, where more classical behavior emerges, such as a collection of many particles, that we should expect a system to be capable of more robust forms of communication.

\section*{Acknowledgements} Sincere thanks to Brenda Rubenstein for pointing us toward various practical applications of two-level systems. 

\newpage
\section*{Appendix}
% \documentclass[conference]{IEEEtran}

% \usepackage{xcolor}
% \usepackage{graphicx} 

% % *** MATH PACKAGES ***
% \usepackage{amsmath}
% \usepackage{amsfonts}

% \begin{document}

% \title{Appendix for \textit{Quantum Analogues for Two Simple Classical Channels}}

% \author{\IEEEauthorblockN{Miles Miller-Dickson}
% \IEEEauthorblockA{School of Engineering,\\
% Brown University\\
% Providence, RI 02912\\
% {\small miles\_miller-dickson@brown.edu}}
% \and
% \IEEEauthorblockN{Christopher Rose}
% \IEEEauthorblockA{School of Engineering,\\
% Brown University\\
% Providence, RI 02912\\
% {\small christopher\_rose@brown.edu}}
% }

% % use for special paper notices
% %\IEEEspecialpapernotice{(Invited Paper)}
% \newcommand{\cnote}[1]{{\bf \boldmath \color{red}  {Chris: \em #1}}}
% \newcommand{\mnote}[1]{{\bf \boldmath \color{blue} {Miles: \em #1}}}
% \newcommand{\be}{\begin{equation}}
% \newcommand{\ee}{\end{equation}}

% \newcommand{\equat}[1]{equation (\ref{eq:#1})}
% \newcommand{\Equat}[1]{Equation (\ref{eq:#1})}

% \newcommand{\xv}{{\bf x}}
% \newcommand{\yv}{{\bf y}}
% \newcommand{\Xmat}{{\bf X}}
% \newcommand{\Ymat}{{\bf Y}}

% \maketitle


\begin{abstract}
Here we include some derivations that were not in the main text. We also include a primer on quantum mechanics for those who are less familiar with the subject.
\end{abstract}

\section*{Quantum mechanics 101}
In quantum mechanics, states are represented by vectors in a complex Hilbert space $\mathcal{H}$, meaning vector coefficients take values in $\mathbb{C}$ and that an inner product has been defined \cite{Sakurai}. Physicists use the notation $|\psi\rangle\in\mathcal{H}$ to refer to these vectors, and will use $\langle \phi|\psi\rangle$ to denote the inner product between two states $|\psi\rangle$ and $|\phi\rangle$. In many applications of quantum information the Hilbert space is just $\mathcal{H}=\mathbb{C}^N$ for some $N\ge2$. In which case the standard basis is sometimes denoted, \cite{Sakurai,Nielsen}

\begin{align*}
    |0\rangle &= (1, 0, ..., 0)^T \\
    |1\rangle &= (0, 1, ..., 0)^T \\
    \vdots\\
    |N-1\rangle &= (0, 0, ..., 1)^T
\end{align*}

And if we write $|\psi\rangle=\sum_{m=0}^{N-1}a_m|m\rangle$ and $|\phi\rangle=\sum_{m=0}^{N-1}b_m|m\rangle$, then the inner product is given by,
\be
    \langle\phi|\psi\rangle = \sum_{m=0}^{N-1} b_m^*a_m
\ee
where $b_m^*$ refers to the complex conjugate of $b_m$, and we have used $\langle m|n\rangle=\delta_{mn}$. 

In section IV we consider an example with $N=2$. In section III, however, when considering the analogue of the Gaussian channel, $\mathcal{H}$ is an infinite dimensional function space, which is appropriate for continuous domains such as the position of a particle. 

Observable quantities, such as position and energy, are represented by Hermitian operators on the Hilbert space. Specifically, the set of eigenvalues of such an operator is precisely the set of possible measurement outcomes for that observable. Hermiticity guarantees that the eigenvalues are real-valued, as expected for measurable quantities. One can expand an arbitrary state in terms of the eigenvectors of any of these Hermitian operators. Most importantly, the \textit{squared modulus} of each coefficient in this expansion has the interpretation of the probability (or probability density) that a measurement of the system will yield the eigenvalue associated to that eigenvector. Thus, eigenvectors are interpreted as states with \textit{definite} observable values, whereas a superposition of eigenvectors cannot be associated with a definite value of that observable (although it \textit{may} correspond to a definite value of some other observable). 

For the purposes of this discussion, we will need to evaluate the time evolution of quantum states. This is determined by the Schrodinger equation, which says,
\be
    i\hbar\frac{d}{dt}|\psi\rangle = \hat H |\psi\rangle
\ee
where $\hat H$ is the operator associated to the energy values, called the \textit{Hamiltonian}, $i=\sqrt{-1}$ is the imaginary unit and $\hbar$ is the so-called \textit{reduced} Planck's constant, whose value is approximately $\hbar\approx 1.05457 \times 10^{-34}$ joule-seconds. A general solution of the time-evolved state, for a \textit{time-independent} energy operator $\hat H$, $|\psi_t\rangle$ is given by,
\be
    |\psi_t\rangle = e^{-i\hat Ht/\hbar}|\psi_0\rangle
\ee
where $|\psi_0\rangle$ represents the initial state of the particle. For our purposes, this will be the state that the sender, Alice, sends to the receiver, Bob. 

We can expand an arbitrary state $|\psi\rangle$ in a basis of eigenvectors of the Hamiltonian $\hat H$, and we will generically label the eigenvectors (sometimes called \textit{eigenstates}) by their eigenvalue, in this case the energy, $|E\rangle$. That is, $\hat H|E\rangle = E |E\rangle$. This notation is ambiguous when there is degeneracy in the eigenspace, but there will be no such degeneracy in our examples and so we will continue to label eigenvectors by their eigenvalue. We also assume that eigenvectors are normalized to unity, so that in particular $\langle E | E \rangle = 1$ and that $\langle E | E' \rangle = 0$ when $E\ne E'$. Writing an arbitrary state as $|\psi\rangle=|\psi_0\rangle=\sum_E a_E |E\rangle$, in the so-called \textit{energy basis}, we see that the Schrodinger equation implies, 
\begin{align*}
    |\psi_t\rangle &= e^{-i\hat Ht/\hbar}\sum_Ea_E|E\rangle \\
    &= \sum_E a_Ee^{-iEt/\hbar}|E\rangle
\end{align*}
since $(\hat H|E\rangle = E |E\rangle)\Rightarrow(e^{-i\hat Ht/\hbar}|E\rangle = e^{-iEt/\hbar} |E\rangle)$. Thus, time evolution is equivalent to scaling the energy eigenvectors by phase factors. In what follows, we will posit a Hamiltonian that models the situation and will expand the initial state built by Alice in terms of the corresponding energy eigenvectors. Then we will insert the appropriate phase factors to get the state that Bob encounters some time $t$ later.

In the case of infinite dimensional Hilbert spaces, such as the one associated with the position of a particle along an axis, states are usually expressed in terms of the so-called ``position basis" $\{|x\rangle\}_{x\in\mathbb{R}}$, where $|x\rangle$ is an eigenstate of the position operator $\hat x$, i.e. $\hat x|x\rangle = x |x\rangle$, meaning that it represents a particle with a definite position $x\in\mathbb{R}$. In this case, one often works with the projection of the state in this basis, denoted $\psi(x) = \langle x | \psi\rangle$, i.e. the component of $|\psi\rangle$ along the basis vector $|x\rangle$. This is usually called the \textit{wavefunction} and we will sometimes include the time dependence of this state and write $\psi(x,t)= \langle x | \psi_t\rangle$. In this basis, the operator corresponding to the momentum of the particle, denoted $\hat p$, turns out to be proportional to the derivative with respect to $x$,
\be
    \hat p = -i\hbar \frac{\partial}{\partial x},
\ee
meaning that the effect of the momentum operator is take the derivative of the wavefunction with respect to position. We will make use of this form of the operator in \equat{freeH} in the next section. 

\section*{Derivation of the time evolution for the Quantum Gaussian Channel}

Computing the time evolution in quantum mechanics is equivalent to scaling the energy eigenfunctions that comprise $\psi(x)$ by the phase factor, $e^{-iE  t/\hbar}$, where $E$ is the energy eigenvalue associated to an energy eigenfunction. In the case of a non-relativistic free particle, the energy operator (in the position basis) is given by, 

\be
    \hat H = \frac{\hat p^2}{2m} = -\frac{\hbar^2}{2m}\frac{\partial^2}{\partial x^2}.
    \label{eq:freeH}
\ee

Its eigenfunctions, $\phi_E(x)=\langle x | E\rangle$, indexed by the energy, satisfy,

\be
    -\frac{\hbar^2}{2m}\frac{\partial^2}{\partial x^2}\phi_E(x) = E \phi_E(x),
\ee

meaning that the eigenfunctions can be written as,

\be
    \phi_E(x) = \frac{1}{\sqrt{2\pi}}e^{i k x}
\ee

where $k = \pm\sqrt{2mE}/\hbar$. Notice that these are in fact the same eigenfunctions for the momentum operator $\hat p = -i\hbar\partial/\partial x$ (and nothing more than Fourier modes). This also follows from the fact that the momentum commutes with the Hamiltonian: $[\hat H, \hat p]=\hat H \hat p-\hat p\hat H=0$, since $\hat H$ is only a function of $\hat p$. The interpretation is that momentum is a conserved quantity (net force $= 0$), whereas position is not ($[\hat H,\hat x]\neq 0$).

Time evolution of each of these modes is given simply by, 
\be
    \phi_E(x) \xrightarrow[]{t} \phi_E(x)e^{-iEt/\hbar}= \frac{1}{\sqrt{2\pi}}e^{i k x - iE t/\hbar}.
    \label{eq:gEnergy}
\ee
\subsection{Energy eigenstates}
It remains to decompose $\psi(x)$ into its energy eigenfunctions (Fourier modes). The Fourier transform of Alice's initial wavefunction $\psi(x) = \langle x | \psi_0 \rangle$, which we'll denote by $\Tilde{\psi}(k)=\langle k | \psi_0 \rangle$, is given by:
\be
    \Tilde{\psi}(k)=\int_{-\infty}^\infty dx \frac{e^{-ikx}}{\sqrt{2\pi}}\frac{e^{-(x - x_0)^2 / 4\sigma_A^2}}{(2\pi\sigma_A^2)^{1/4}} 
\ee

We can make use of the following special integral,
\be
    \int_{-\infty}^\infty dz e^{\frac{1}{2}iaz^2 + iJz} = \sqrt{\frac{2\pi i}{a}}e^{-iJ^2/2a}
\ee
Setting $z = x-x_0$ in the $\Tilde{\psi}(k)$ integral leaves, 
\be
    \Tilde{\psi}(k)=\frac{e^{-ikx_0}}{\sqrt{2\pi}(2\pi\sigma_A^2)^{1/4}}\int_{-\infty}^\infty dz e^{-ikz}e^{-z^2 / 4\sigma_A^2} 
\ee
from which we see we can make use of our special integral by setting $a=i/2\sigma_A^2$ and $J=-k$. Thus, we have, 
\be
    \Tilde{\psi}(k)=\frac{e^{-ikx_0}\sqrt{2\sigma_A}}{(2\pi)^{1/4}}e^{-k^2\sigma_A^2}.
\ee
Hence, the time evolved state $\psi(x,t) = \langle x|\psi_t\rangle$ can be written as,
\begin{align*}
    \psi(x,t) &= \int_{-\infty}^\infty dk \Tilde{\psi(k)}\frac{1}{\sqrt{2\pi}}e^{i k x - iE t/\hbar} \\ &=\frac{\sqrt{2\sigma_A}}{\sqrt{2\pi}(2\pi)^{1/4}}\int_{-\infty}^\infty dk e^{-k^2\sigma_A^2 + i k (x-x_0) - iE t/\hbar}.
\end{align*}

Recall that $k$ was defined by $k = \pm\sqrt{2mE}/\hbar$, and so $E = \hbar^2 k^2/2m$. We see we can again make use of our special integral, this time setting $a = 2i\left(\sigma_A^2 + \frac{i\hbar t}{2m}\right)$, and $J = x-x_0$.
And so we have, 
\begin{align*}
    \psi(x,t)&=\frac{\sqrt{2\sigma_A}}{\sqrt{2\pi}(2\pi)^{1/4}}\sqrt{\frac{\pi}{\sigma_A^2 + \frac{i\hbar t}{2m}}}\exp\left(\frac{-(x-x_0)^2}{4\left(\sigma_A^2 + \frac{i\hbar t}{2m}\right)}\right)\\
    &=\frac{1}{\sqrt{\sqrt{2\pi}\left(\sigma_A + \frac{i\hbar t}{2m\sigma_A}\right)}}\exp\left(\frac{-(x-x_0)^2}{4\left(\sigma_A^2 + \frac{i\hbar t}{2m}\right)}\right).
\end{align*}
\subsection{Energy considerations for the Gaussian channel}
One other consideration is that of the finite amount of time taken for Alice to prepare her state. Doing so more rapidly will incur an energetic cost. Since the particle initially emerges at $x=0$, the energy required for particle placement is proportional to the energy necessary to move a particle from $x=0$ to some $X = x_n$ in a specified amount of time $T$, where $n$ indexes the placement run.  Thus, the minimum average power $P$ expended to place particles goes as,
\be
\label{eq:Power}
P
=
\frac{1}{2(T+t)} m \mathbb{E} \left [ \left (\frac{X}{T} \right )^2 \right ]
\ee
where $\frac{1}{T+t}$ is the rate at which the communication scheme cycles (particles are placed then measured), $m$ is the particle mass and the $\mathbb{E}[\cdot]$ denotes the expected value over particle placement positions $X$. For notational simplicity we will write \equat{Power} as,
\be
\label{eq:Power2}
P
=
\beta \mathbb{E} \left [ X^2 \right ]
\ee
where
$$
\beta = \frac{m}{2T^2(T+t)}
$$

\section*{Derivation of the time evolution for the two-level channel}
We model transitions between the two levels using a parameter $\epsilon\in\mathbb{R}$ in a Hamiltonian of the form,
\be
    \hat H = 
    \begin{pmatrix}
        E & \epsilon \\
        \epsilon & E+\Delta
    \end{pmatrix} .
\ee

Here we assume Alice can set her initial state as,
\be
    |\psi_0\rangle=\sqrt{1-p}|0\rangle + \sqrt{p}|1\rangle
    \label{eq:2levelState}
\ee
either with $p$ close to $0$ (so that $|\psi_0\rangle$ is close to $|0\rangle$) or with $p$ close to $1$ (so that $|\psi_0\rangle$ is close to $|1\rangle$). Here $p$ gives the initial probability of a measurement resulting in the state $|1\rangle$, and so $1-p$ is the probability of measuring the state $|0\rangle$. Hence, we can use $p(1-p)$ as a measure of the variance for the corresponding distribution, $|\langle z | \psi_0\rangle|^2$, where $z\in\{0,1\}$ (Bernoulli distribution).

We could also include a relative phase factor between the $|0\rangle$ and $|1\rangle$ states in Alice's state $|\psi_0\rangle$, such as $e^{i\theta}$ for some $\theta\in\mathbb{R}$, but since Bob only makes measurements along the $\{|0\rangle,|1\rangle\}$ basis, any such phase factor will drop out of the discussion when taking the modulus squared.

For arbitrary $\epsilon\ge0$ the energy eigenstates can be computed (the $\epsilon<0$ case can be treated similarly). We will label them by their energy eigenvalues, $E_\pm$, 
\be
    |E_\pm\rangle = 
    \frac{1}{\sqrt{2a}}\begin{pmatrix}
        \sqrt{a\mp b} \\
        \pm\sqrt{a\pm b}
    \end{pmatrix}
\ee
where,
\begin{align*}
    a &= \frac{\sqrt{\Delta^2 + 4\epsilon^2}}{2}, \\
    b &= \frac{\Delta}{2}
\end{align*}
and, 
\begin{align}
    E_\pm &= E + \frac{\Delta\pm\sqrt{\Delta^2+4\epsilon^2}}{2} \\
    &= E \pm a + b,
    \label{eq:Energy}
\end{align}
as can be easily verified. 

We can write Alice's starting states, $|0\rangle$ and $|1\rangle$, in terms of these energy states.
\begin{align*}
    |0\rangle &= \begin{pmatrix}
        1 \\
        0
    \end{pmatrix} = \sqrt{\frac{a-b}{2a}}|E_+\rangle + \sqrt{\frac{a+b}{2a}}|E_-\rangle,\\
    |1\rangle &= \begin{pmatrix}
        0 \\
        1
    \end{pmatrix} = \sqrt{\frac{a+b}{2a}}|E_+\rangle - \sqrt{\frac{a-b}{2a}}|E_-\rangle.
\end{align*}
As before, time evolution is given by simply scaling the energy states by a phase factor, in this case,
\be
    |E_\pm\rangle \xrightarrow[]{t} |E_\pm\rangle e^{-iE_\pm t/\hbar}
\ee
    just as in \equat{gEnergy}. 

Writing $|\psi_0\rangle$ in terms of $|E_\pm\rangle$,
\begin{align*}
    |\psi_0\rangle = \frac{1}{\sqrt{2a}}\left(\sqrt{(1-p)(a-b)} + \sqrt{p(a+b)}\right)|E_+\rangle \\
    + \frac{1}{\sqrt{2a}}\left(\sqrt{(1-p)(a+b)} - \sqrt{p(a-b)}\right)|E_-\rangle.
\end{align*}
And so the time evolved state is given by,
\begin{align*}
    |\psi_t\rangle &= \frac{1}{\sqrt{2a}}\left(\sqrt{(1-p)(a-b)} + \sqrt{p(a+b)}\right)|E_+\rangle e^{-iE_+t/\hbar} \\
    &+ \frac{1}{\sqrt{2a}}\left(\sqrt{(1-p)(a+b)} - \sqrt{p(a-b)}\right)|E_-\rangle e^{-iE_-t/\hbar} \\
    &= \frac{e^{-i(E+b)t/\hbar}}{a} \\
    &\times
    \begin{pmatrix}
        a\sqrt{1-p}\cos(\frac{at}{\hbar}) + i\left(b\sqrt{1-p} - \epsilon\sqrt{p}\right)\sin(\frac{at}{\hbar}) \vspace{0.1in}\\
        a\sqrt{p}\cos(\frac{at}{\hbar}) - i\left(\epsilon\sqrt{1-p} + b\sqrt{p}\right)\sin(\frac{at}{\hbar})
    \end{pmatrix},
\end{align*}
where we used $\epsilon^2 = a^2 - b^2$.

Lastly, we compute the modulus squared of each coefficient to give the transition probabilities,
\be
    \label{eq:tProb2level}
    |\langle z | \psi_t\rangle|^2 = \frac{1}{a^2}
    \begin{pmatrix}
        \zeta_p\cos(\frac{2at}{\hbar}) + a^2(1-p)-\zeta_p  \\
        - \zeta_p\cos(\frac{2at}{\hbar}) + a^2p+\zeta_p
    \end{pmatrix}_z,
\ee
where, 
\be
    \zeta_p = \frac{\epsilon}{2}\left(\epsilon(1-2p)+2 b\sqrt{p(1-p)}\right)
\ee
with $z\in\{0,1\}$.

% \newpage
\bibliographystyle{plain}
\bibliography{refs}


\newpage
\bibliographystyle{plain}
\bibliography{refs}

\end{document}